\documentclass[apj, iop]{emulateapj}

\newcommand{\msun}{\ensuremath{{{M}}_{\scriptscriptstyle \odot}}}
\newcommand{\mhalo}{\ensuremath{M_h}}

\newcommand{\msigma}{\ensuremath{M_\mathrm{BH}}--\ensuremath{\sigma}}
\newcommand{\mvc}{\ensuremath{M_\mathrm{BH}}--\ensuremath{V_c}}
\newcommand{\kms}      {\ensuremath{~\mathrm{km~s^{-1}}}} 
\newcommand{\mbh}{\ensuremath{M_\mathrm{BH}}}

\newcommand{\beq}{\begin{equation}}
\newcommand{\eeq}{\end{equation}}
\def\Omm{{\Omega_m}}
\def\Ommz{{\Omega_m^{\,z}}}

\def\Omk{{\Omega_k}}
\def\Oml{{\Omega_{\Lambda}}}

\begin{document}

\title{How important is the dark matter halo for black hole growth?}

\author{Marta Volonteri$^{1}$,  Priyamvada Natarajan$^{2,3}$ \& Kayhan Gultekin$^{1}$ }
\affil{1 Department of Astronomy, University of Michigan, Ann Arbor, MI, USA}
\affil{2 Department of Astronomy, Yale University, New Haven, CT, USA}
\affil{3 Institute for Theory and Computation, Harvard-Smithsonian Center for Astrophysics, 60 Garden Street, Cambridge, MA, USA}

\begin{abstract}
In this paper, we examine if the properties of central black holes in galactic nuclei 
correlate with their host dark matter halo. We analyze the entire sample 
of galaxies where black hole mass, velocity dispersion $\sigma$, and asymptotic circular 
velocity $V_c$ have  all been measured.  We fit \msigma\ and \mvc\  
to a power law, and find that in both relationships the scatter and slope are similar. This model-independent analysis suggests 
that although the black hole masses are not uniquely determined by dark matter halo mass, when considered  for the  current sample as a whole, the \mvc\ 
correlation may be as strong (or as weak) as $M_\mathrm{BH}$--$\sigma$. Although the data are sparse, there appears to be more scatter in the correlation 
for both $\sigma$ and $V_c$ at the low--mass end. This is not unexpected given our current 
understanding of galaxy and black hole assembly. In fact, there are several compelling reasons 
that account for this:  (i)  SMBH formation is likely less efficient in low-mass 
galaxies with large angular momentum content; (ii) SMBH growth is less efficient in low-mass disk galaxies 
that have not experienced major mergers; (iii) dynamical effects, such as gravitational recoil, increase 
scatter preferentially at the low-mass end. Therefore, the recent observational claim of the absence of 
central SMBHs in bulgeless,  low mass  galaxies, or deviations from the correlations defined by high-mass 
black holes in large galaxies today is, in fact, predicated by current models of black hole growth. We show 
how this arises as a direct consequence of the coupling between dark matter halos and central black holes at 
the earliest epochs.
\end{abstract}


\section{Introduction}

Demography of local galaxies suggests that most galaxies
harbor a quiescent super-massive black hole (SMBH) in their nucleus today. Observational evidence 
also points to the  existence of a strong correlation between the mass of the central SMBH  and the properties 
of the host spheroid (Ferrarese \& Merritt 2000, Gebhardt et al. 2000; 
Tremaine et al. 2002; Marconi \& Hunt 2003; H\"aring \& Rix 2004; G\"ultekin et al. 2009) and possibly the host
halo (Ferrarese 2002) in nearby galaxies. These correlations are strongly 
suggestive of co-eval growth of the SMBH and the stellar component likely via 
regulation of the gas supply in galactic nuclei from the earliest times 
(Haehnelt, Natarajan, Rees 1998; Silk \& Rees 1998; Kauffmann \& Haehnelt 2000; 
Fabian 2002; King 2003; Thompson, Quataert \& Murray 2005; Natarajan \& Treister 2009, Booth \& Schaye 2010). 

In a recent pair of papers  \cite[KBC and KB thereafter respectively]{Kormendy2011a,Kormendy2011b} have argued that 
(i) SMBH masses appear not to correlate with galaxy disks, and 
(ii)  it is the morphology of the host galaxy that determines whether a correlation exists or not
with the SMBH, and the correlation is not contingent on properties of the dark matter halo. 
In this letter, we re-examine the evidence and demonstrate that the properties of the dark
matter halo are, in fact, relevant to the growth and assembly of the central
SMBH. Furthermore, we show that correlations between the mass of the 
SMBH and the host dark matter halo are inevitable during hierarchical growth within 
the $\Lambda$CDM model wherein galaxy assembly is
driven by the mergers of dark matter halos \citep[see also][]{Peng2007,Jahnke2010}. Our goal here is to clarify that some very plausible 
models for the formation of SMBH seeds at the earliest cosmic epochs do indeed correlate key 
dark matter halo properties to the mass of the assembling SMBH. In addition these same  models do predict 
that  bulgeless galaxies today are frequently bereft of SMBHs as found by KB 
(Volonteri,  Lodato \& Natarajan 2008; Volonteri \& Natarajan 2009).
Our interpretation is that at earlier times the properties of the assembling SMBH are more
tightly coupled to properties of the dark matter halo as its growth
is driven by the merger history of halos. However, at later times 
the final mass of the SMBH becomes more tightly
coupled to the small scale local baryonic distribution.   

\section{Black holes and Bulges}

In their recent analysis of correlations between black holes and dark
matter halos, KB found that halo mass (measured with circular
velocity, $V_c$) was a far poorer predictor of black-hole mass
compared to, e.g., galaxy stellar velocity dispersion ($\sigma$) or
bulge luminosity ($L_\mathrm{bulge}$), especially for smaller
galaxies.  This trend is further displayed by an apparent break down
in small galaxies of the otherwise tight, linear correlation between
$V_c$ and $\sigma$.  A linear correlation is expected for simple
galaxy models so that the observed deviation in small galaxies is
argued to imply no correlation.  KBC interpret the tight, linear
correlation between $V_c$ and $\sigma$ in large galaxies arising from
the well known ``cosmic conspiracy.''

As per KB it does appear that today $M_\mathrm{BH}$ and $V_c$ are, in
fact, correlated but possibly more weakly so, particularly at lower
masses, compared to the $M_\mathrm{BH}$--$\sigma$ and
$M_\mathrm{BH}$--$L_\mathrm{bulge}$ relations. However, we offer an
alternative interpretation for why this is so.  We argue here that
this is likely a direct consequence of the role of merging dark matter
halos that drive SMBH growth.  Our view hinges on the fact that the
coupling between SMBH growth and dark matter was necessarily strong at
high redshifts as merger rates determine the assembly history, and
major mergers that trigger accretion episodes are more frequent at
high redshift for the most massive halos.  At late times, however, the
SMBH mass itself is more tightly coupled to the properties of the
baryonic galactic nucleus, in particular, for low mass galaxies that
have experienced practically no major mergers in their entire
lifetime.  The key point here is that major mergers trigger
simultaneous SMBH growth and star formation causing a tight coupling
between these two components in the galactic nucleus only for massive
halos.

To explore this interpretation further, we first re-examine the data
from KBC from a purely empirical perspective.  The sample is composed
of 25 galaxies with dynamical measurements of the black hole mass,
$M_\mathrm{BH}$, and high-quality measurements of velocity dispersion,
$\sigma$, and asymptotic circular velocity, $V_c$ (Fig.~\ref{data}).
Galaxy properties are listed in Table 1 of the Supplementary
Information in KBC.  We fit the data with a functional form of
\begin{equation}
\log_{10}\left(\frac{M_\mathrm{BH}}{\msun}\right) = A + B \log_{10}\left( \frac{V_c}{200\kms} \right).
\label{eq:mvc}
\end{equation}
Using a symmetric least-squares fit, we find $A = 7.2 \pm 0.05$ and $B
= 7.60 \pm 0.40$ with $\chi^2/\textrm{d.o.f.} = 7.2$, indicating that
the data are very unlikely to have come from the model used, in strong
agreement with KB.  

We then expand the model to include a log-normal
scatter about the relation of standard deviation $s_0$, The inclusion
of a scatter term of some sort is essential since the deviations from
a log-linear relation are in excess of the measurement errors
\citep{Hogg2010}.  We fit using the methods of \cite{Gultekin2009}, a
generalized maximum likelihood method that can handle measurement
errors in the independent and dependent variables (assumed Gaussian in
log space) as well as upper limits.  We find $A = 7.39 \pm 0.14$, $B =
4.22 \pm 0.93$, and $s_0 = 0.53 \pm 0.10$.  So this indicates that a
correlation can be inferred from the data. Compared to the \msigma\ relation, the scatter in the
\mvc\ relation is (\emph{i}) slightly larger for this sample, for
which we find $A = 8.06 \pm 0.14$, $B = 3.95 \pm 0.72$, and $s_0 =
0.50 \pm 0.09$; (\emph{ii}) significantly larger than the entire
sample in \cite{Gultekin2009} ($s_0=0.44\pm0.06$) and (\emph{iii})
much larger than the elliptical-only sample in \cite{Gultekin2009}
($s_0 = 0.31 \pm 0.06$).

Given that the sample was selected based on the ability to measure
$V_c$ in each galaxy, the actual scatter in the \mvc\ relation may be
even larger since those galaxies in which $V_c$ is difficult to
measure will tend to be outliers.  The fact that the inferred scatter
in the \mvc\ relation is not smaller than the scatter in the \msigma\
implies that the halo mass is not driving the correlation at a higher 
level than the physical process that sets the bulge properties. This
first assessment  corroborates one of the main conclusions of KB, that
central black hole's mass today is not {\it uniquely} determined by the mass of the dark matter halo. 
The similar level of scatter in  the \msigma\  and \mvc\ fits, however, confirms that there 
is a trend in the entire \mvc\ sample (plus notable outliers; see Fig.\ \ref{data}.  For
example, there is only one galaxy with $V_c > 250\kms$ and $\mbh <
10^8\ \msun$ and only one with $V_c < 200\kms$ and $\mbh >
2\times10^7\ \msun$.  It is also possible to interpret the data as having only a weak correlation below
$\mbh < 5 \times 10^7\ \msun$).  We show below how this trend is expected to naturally
arise in a set of physically well motivated models for the formation
of SMBH seeds at the earliest epochs without requiring black holes to
partake in exotic nonbaryonic physics.

\begin{figure}
\includegraphics[width= \columnwidth]{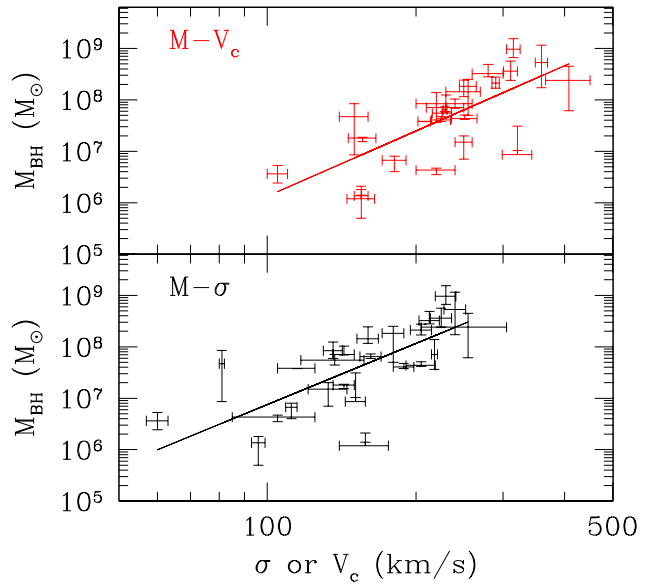}
\caption{Top panel: \mvc\ relation for the 25 galaxies in Table 1of KBC. Bottom panel: \msigma\ relation from Kormendy's table. The best fit in each case is also shown.}
\label{data}
\end{figure}

\section{Links between SMBH formation, halo mass and spin}

We focus on a class of SMBH seeding models to illustrate that the observed lack of
central black holes in low velocity dispersion bulgeless galaxies
 today need not imply a lack of correlation between SMBH and dark
matter halo properties at earlier epochs \citep{VLN2008,VN09}. We describe below a class of black hole seed 
formation models where the seed properties are initially
correlated to the host dark matter properties at high redshifts. 
Evolving such models via the merger driven accretion prescription
over cosmic time, Volonteri \& Natarajan (2009) find that a key prediction is that {\it low mass 
bulgeless galaxies today are unlikely to host nuclear black holes}.
The relevant host dark matter halo property in this picture is the spin. In a physically motivated model for the 
formation of heavy SMBH seeds (in contrast to the lower mass remnant
seeds from Population III stars)  according to the prescription
described in Lodato \& Natarajan (2006; 2007) there is a limited
range of halo spins and halo masses that are viable sites for the
formation of seeds. Details of these models have been presented in
several published papers and here we merely paraphrase the relevant details.

In direct collapse models, SMBH seeds can form at high redshift ($z>15$) in metal-free galaxies with  $T_{\rm vir}>10^4$ K, where atomic hydrogen cooling becomes effective \citep{Koushiappas2004, BVR2006, LN2006, LN2007}.  Here we refer to
\citet{LN2006,LN2007}, for more details of the seeding model, wherein the
development of non-axisymmetric spiral structures drives mass infall
and accumulation in a pre-galactic disk with primordial
composition. The mass accumulated in the center of the halo (which
provides an upper limit to the SMBH seed mass) is given by:
\begin{equation}
M_{\rm BH}= m_{\rm d}\mhalo\left[1-\sqrt{\frac{8\lambda}{m_{\rm d}Q_{\rm c}}\left(\frac{j_{\rm d}}{m_{\rm d}}\right)\left(\frac{T_{\rm gas}}{T_{\rm vir}}\right)^{1/2}}\ \right] 
\label{mbh}
\end{equation}
for 
\begin{equation}
\lambda<\lambda_{\rm max}=m_{\rm d}Q_{\rm c}/8(m_{\rm d}/j_{\rm d}) (T_{\rm
  vir}/T_{\rm gas})^{1/2}
\label{lambdamax} 
\end{equation}
and $M_{\rm BH}=0$ otherwise. Here $\lambda_{\rm max}$ is the maximum
halo spin parameter for which the disk is gravitationally unstable,
$m_d$ is the gas fraction that participates in the infall, $j_{\rm d}$ is its specific angular momentum, and $Q_{\rm
c} \gtrsim 1$ is the Toomre parameter.  The efficiency of the seed assembly process ceases at large halo
masses, where the disk undergoes fragmentation instead. This occurs
when the virial temperature exceeds a critical value $T_{\rm max}$,
given by:
\begin{equation}
\frac{T_{\rm max}}{T_{\rm gas}}=\left(\frac{4\alpha_{\rm c}}{m_{\rm
d}}\frac{1}{1+M_{\rm BH}/m_{\rm d}\mhalo}\right)^{2/3},
\label {frag}
\end{equation}
where $\alpha_{\rm c}\approx 0.06$ is a dimensionless parameter measuring the
critical gravitational torque above which the disk fragments \citep{RLA05}.

\begin{figure}
\includegraphics[width= \columnwidth]{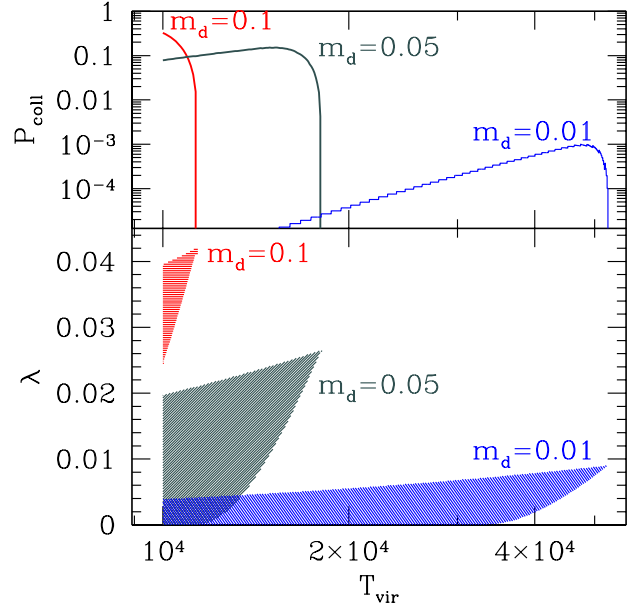}
\caption{Parameter space (virial temperature, spin parameter) for
  SMBH formation. Here we select halos with $T_{\rm vir}>10^4$ K
  at $z=15$ for different gas fractions that participate in the infall ($m_{\rm d}$). Other adopted parameters are $m_{\rm d}=j_{\rm d}$,
$\alpha_{\rm c}=0.06$, and here we consider the $Q_{\rm c}=2$ case (cf. Volonteri et al. 2008).
The gas has a temperature $T_{\rm gas}=5000$ K.
  The shaded areas in the bottom panel show the range of virial temperatures and spin parameters where disks are
  Toomre unstable and the joint conditions, $\lambda<\lambda_{\rm max}$
  (Equation~\ref{lambdamax}) and $T_{\rm vir}<T_{\rm max}$ (Equation~\ref{frag}, providing the minimum spin parameter, $\lambda_{\rm min}$ value below which the disk is globally prone to fragmentation) are fulfilled. 
  The top panel shows the probability of SMBH formation, by integrating the lognormal distribution of spin parameters between $\lambda_{\rm min}$ and  $\lambda_{\rm max}$.}
\label{Pcoll_lambda}
\end{figure}

To summarize, a dark matter halo is characterized by its mass \mhalo\ 
(or virial temperature $T_{\rm vir}$) and by its spin parameter
$\lambda$.  If $\lambda_{\rm min}<\lambda<\lambda_{\rm max}$ (see eqn.~\ref{lambdamax} and~\ref{frag}) and $10^4$ K$<T_{\rm vir}<T_{\rm max}$ (eqn.~\ref{frag}), then a seed SMBH can form in the center. 
Hence SMBHs form (i) only in halos within a given range of virial temperatures, hence, halo masses, and (ii) only within a narrow range of  spin parameters, as shown in Figure~\ref{Pcoll_lambda}. 
Low mass halos ($T_{\rm vir}<10^4$ K) and halos with high values of $\lambda$, most likely leading to disk-dominated galaxies, are strongly disfavored as SMBH seed sites in this model or in others that rely on global dynamical instabilities  \citep[see][]{BVR2006,BV2010}. Therefore, these models of SMBH formation naturally predict that bulgeless galaxies are unlikely to host SMBHs today, in agreement with KBC conclusions.  

\section{Link between SMBH growth and halo mass}

We briefly outline familiar aspects of the mass assembly history of black holes. Details and discussion on the relevant literature and background can be found in  Volonteri \& Natarajan (2009) and references therein. 

Accretion episodes are triggered by major mergers, and they drive the mass growth of the central black holes. These very major mergers are also implicated in the formation of  bulges. In the $\Lambda$CDM framework, the lowest mass halos, though, experience mostly minor mergers, which do not trigger accretion episodes and hence do not grow the SMBH, nor form a bulge. Effectively, for the lowest mass halos, growth of the stellar  component of the galaxy  and the central SMBH are not coeval, given that major mergers are the key drivers. Even if these low mass halos are initially seeded we predict that most low surface brightness, low-mass bulgeless galaxies at the present time  do not host central black holes (Volonteri \& Natarajan 2009). In galaxies that are now high mass, SMBH growth and assembly of the stellar component most likely proceeded co-evally over cosmic history.  Regulation of the gas physics that underlies the evolution of both processes occurred in the inner regions of the galaxies, triggered by galaxy mergers. 

In currently successful theoretical models that fit observational data of accreting black holes over a range of redshifts, the growth of a SMBH is modulated by the history of its host, which in turn depends on a merger history that is set by the properties of the dark matter halo. The host dark matter halo and its merger properties in the context of $\Lambda$CDM are crucial to black hole growth at early epochs. The results and predictions of these models have been presented in detail in several papers as mentioned above and there is a fair degree of consensus amongst the community that these analytic Monte-Carlo merger tree based models do adequately capture most of the relevant physics.

In order to clarify the relevant aspects of these models for our discussion here, we now explicitly develop an {\it analytical model} that uses only halo merger rates to delineate the role of dark matter halos for black hole growth. As an ansatz, we start with the premise that SMBHs grow ``healthily" only if their host experiences a major merger. Furthermore, we assume major mergers bring a SMBH on the scaling relations expected for their hosts, for instance via self-regulation and AGN feedback \citep[e.g.,][]{Silk1998,Fabian1999,DiMatteo2005}. The details of these so-called AGN feedback processes are unsettled at the present time, as there is no entirely convincing theoretical model or simulation for that matter wherein the detailed microphysics of these feedback processes is captured. For our purposes, however,  it is sufficient to simply invoke them as a means to setting up the scaling relations.  The existence of these scaling relations suggests that such processes ought to occur. 

Here, we use {\it just} the halo merger rate as the starting point for evolving black hole masses over time. High resolution dark matter only cosmological simulations provide us the merger rates for halos e.g., \cite{Fakhouri2010}. In fact, a 
simple fitting  formula can be derived (Equation 1 in Fakhouri et al. 2010) for the merger rate per unit redshift and mass ratio ($\xi\le1$) at fixed halo mass\footnote{Halo mass can be translated into virial circular velocity:
\begin{equation}
 V_{\rm c}= 142\kms \left[\frac{\mhalo}{10^{12}\ \msun }\right]^{1/3} 
\left[\frac {\Omm}{\Ommz}\ \frac{\Delta_{\rm c}} {18\pi^2}\right]^{1/6} 
(1+z)^{1/2},  
\end{equation}
where $\Delta_{\rm c}$ is the over-density relative
to the critical density.  For a WMAP5 cosmology we adopt the fitting formula $\Delta_{\rm c}=18\pi^2+82 d-39 d^2$
(Bryan \& Norman 1998), where $d\equiv -1+{\Omm (1+z)^3}/({\Omm
(1+z)^3+\Oml+\Omk (1+z)^2})$.
}:
\beq
\frac{dN_m}{d\xi dz}(\mhalo) = A_0\left(\frac{\mhalo}{10^{12} M_0}\right)^{\alpha}\xi^{\beta}\exp\left[\left(\frac{\xi}{\tilde{\xi}}\right)^{\gamma}\right](1+z)^{\eta}.
\label{mjm}
\eeq
with $A_0=0.0104$, $\alpha=0.133$, $\beta=-1.995$, $\gamma=0.263$, $\eta=0.0993$ and $\tilde{\xi}=9.72\times10^{-3}$. We integrate the merger rate between $z=0$ and $z=3$, and mass ratio $\xi >0.3$ (major mergers. The minimum mass ratio that triggers  growth episodes for the central black hole can be as low as $\xi > 0.1$, however for the successful formation of bulges the threshold needs to be raised, e.g., Somerville et al. 2000).  This gives the number of major mergers a halo of a given mass experiences between $z=0$ and $z=3$.  These major mergers are expected to be responsible for the simultaneous growth of bulges and SMBHS (e.g., di Matteo et al. 2005). If a halo experiences at least one major merger, then its SMBH has a chance of growing in a self-regulated manner (via AGN feedback, as mentioned above) with its host. If a halo does not experience any major merger, than the growth of the SMBH is expected to be decoupled from the properties of the host, being driven by, e.g., secular effects (cf. KBC). 

It is apparent from Equation~5 that the major merger rate is an increasing function of halo mass/circular velocity.  In fact we  find that the expected number of mergers between $z=0$ and $z=3$ with mass ratio $\xi >0.3$ is approximately 0.4 for $\mhalo=10^8\ \msun$, 0.5 for $\mhalo=10^9\ \msun$, 0.7 for $\mhalo=10^{10}\ \msun$, and it is above unity only for $\mhalo>10^{11}\ \msun$.  Therefore, {\it if the correlation between SMBH mass and hosts is  established by the simultaneous growth of the bulge and SMBH triggered by major mergers, SMBHs in low mass galaxies are simply  more likely to be outliers}. Here, we have shown clearly the role that major mergers play in the triggering of mass growth for central black holes, and that it is the dark matter halo mass that is the key determinant for the deriving the major merger rate. 
Deviations from the scaling relations that are a direct consequence of this merger driven scenario are therefore expected and inevitable for low mass halos.  Below we show explicitly how these deviations are likely set up for low mass halos/galaxies and how these are a result of their growth histories.

To quantify deviations from the \mvc\ relation, we run a Monte Carlo simulation of the growth of 560,000 galaxies, and ask at $z=0$ how many SMBHs are ``ungrown" (no major mergers), ``ejected" (kicked because of gravitational recoil), or ``healthy" (at least one major merger brought the SMBH on the correlation, and if a SMBH has been ejected, a major merger after the ejection is required to bring the SMBH back onto the  \mvc\ relation). 

We create a Monte Carlo realization of halo mergers per halo as a function of mass, where the probability of a major merger and the mass ratio of a given major merger are drawn from the distribution defined in Equation~\ref{mjm}.  We first assume that initially each halo hosts a SMBH, and that the its mass is predicted by the $M_{\rm BH}-V_c$ relation that is  proportional to the halo mass to the 4/3 power (combining the relationships \mvc\ and \mhalo--$V_c$). From the mass ratio of merging halos ($\xi$) we can derive the mass ratio of merging SMBHs ($q=\xi^{4/3}$). We further assume a probability distribution of SMBH spins (we assumed a random distribution between 0 and 1), and of inclination between SMBH spins and orbital angular momentum at the time of SMBH merger \citep[we assumed partial alignment,  cf. the `hot' case in][results are qualitatively insensitive to this choice]{Dotti2010,VGD2010}  and derive gravitational recoil velocities \citep{vanMeter2010} for each SMBH merger in our sample. By comparing these recoil velocities to escape velocities from halos, assumed to be described by NFW profiles, we can derive an ejection probability \citep[see][]{VGD2010}. We then run a second realization where we assume a SMBH ``occupation fraction" ($OF$, i.e., the fraction of halos that hosts SMBHs) derived from Figure 4 in VLN2008 ($Q_c=2$). Using a linear fit in log-space, the occupation fraction scales with the halo mass as:
\beq
OF=\min(10^{1.3\log(M_{\rm h}/10^{12}\msun)},1).
\label{OF}
\eeq

\begin{figure}
\includegraphics[width= \columnwidth]{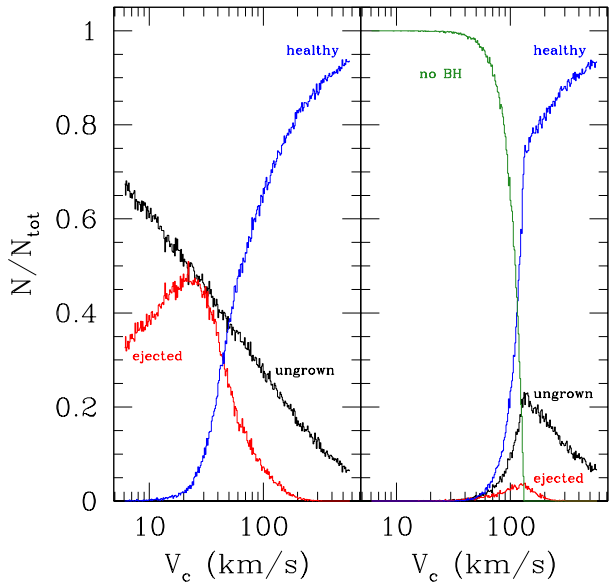}
\caption{Distribution of SMBH growth histories as a function of halo mass. If a SMBH is hosted in a galaxy that has not experienced a major merger ($\xi>0.3$) since $z=3$ we dub it ``ungrown''. If the galaxy experiences a major merger that results in the SMBH being ejected, and no further major merger brings in SMBH, we assume the galaxy has no SMBH (``ejected'').  A ``healthy'' SMBH is likely to sit on or near the expected \mvc\ relation. ``Healthy'' SMBHs are hosted in galaxies where at least one major merger brought the SMBH on the correlation, and, if a SMBH has been ejected, we require a major merger after the fact to bring in a new SMBH and grow it  onto  the \mvc\ relation. The left panel assumes that all halos host a SMBH, regardless of mass. The right panel  assumes an ``occupation fraction" of SMBHs derived from Figure 4 in VLN2008 ($Q_c=2$).}
\label{MCgrowth}
\end{figure}

Figure~\ref{MCgrowth} shows the distribution of SMBH growth histories. It is apparent that most SMBHs in galaxies with $V_c \gtrsim100 \kms$ have experienced al least one major merger between $z=3$ and $z=0$. Given the assumption that major merger-triggered accretion brings SMBHs on or near the expected $M_{\rm BH}-V_c$ relation, these SMBHs are ``healthy".

The fraction of outliers increases as we move towards lower and lower masses, where most SMBHs are either ungrown or ejected, or where the galaxy never hosted a SMBH in the first place. We note that the effect of gravitational recoil (ejections) is strongest at intermediate galaxy masses.  This is because low-mass galaxies have few mergers while high-mass galaxies have large escape velocities. Since low-mass galaxies have a small number of major mergers, or they have no SMBH, even if the ejection probability -- based on the comparison between the escape and recoil velocities -- is close to 100\%, if a galaxy has no major merger at all, then the ejection probability convolved with the merger probability is zero.

This analytical model therefore predicts the halo mass where the transition from healthy growth (triggered by major mergers) to unhealthy SMBHs (in the sense that their masses are not set by merger-driven accretion) occurs. One can integrate Equation 5 at different redshifts to find where this transition occurs at early times, if desired. The transition corresponds to the halo mass (or circular velocity) where the average number of major mergers drops below unity.

\begin{figure}
\includegraphics[width= \columnwidth]{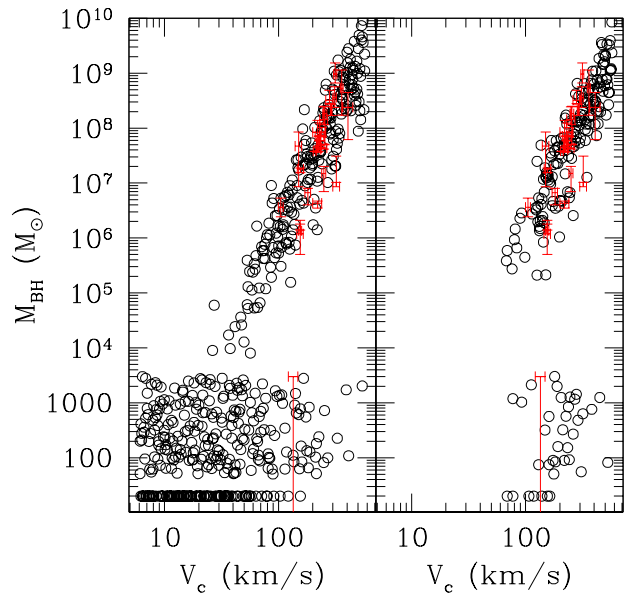}
\caption{Circles: $M_{\rm BH}-V_c$ relation from our MC simulations. Here at every major merger a SMBH increases its mass by $10^8\ \msun (V_{\rm c}/350 \kms)^4\times 10^{0.5\Delta}$ where $\Delta$ is normally distributed about 0 with standard deviation 1. The mass of ``ungrown'' SMBHs is set to $300\msun\times10^{2R-1}$, where R is randomly distributed between 0 and 1. Ejected SMBHs are arbitrarily set to $M_{\rm BH}=10\ \msun$ to show which galaxies are affected. The left panel assumes that all halos host a SMBH, regardless of mass. The right panel  assumes an "occupation fraction" of SMBHs derived from Figure 4 in VLN2008 ($Q_c=2$). Galaxies from KBC are shown with errorbars. M33 is shown as the upper limit at $3000\, \msun$.}
\label{MCgrowth2}
\end{figure}

In Figure~\ref{MCgrowth2} we present the $M_{\rm BH}-V_c$ relation for 560 random galaxies in our Monte Carlo sample (uniformly drawn out of the complete sample of 560,000) and compare it to the $M_{\rm BH}-V_c$ of the 25 galaxies described in Section 2 (from Table 1 in KBC, augmented by the upper limit for M33). Here we assume that at every major merger a SMBH increases its mass by $10^8\ \msun (V_{\rm c}/350 \kms)^4\times 10^{0.5\Delta}$ where $\Delta$ is normally distributed about 0 with standard deviation 1 (see section 2 for details). These are ``healthy" SMBHs. The mass of ``ungrown'' SMBHs is set to $300\msun\times10^{2R-1}$, where R is randomly distributed between 0 and 1. ``Ejected" SMBHs are arbitrarily set to $M_{\rm BH}=10\ \msun$ to demarcate the affected galaxies. The Monte Carlo sample clearly occupies the same area occupied by the real galaxies. Therefore, the deviations from the scaling relation at low masses today reflects the growth history of black holes that is driven principally by the dark matter halo mass.

\section{Discussion and Conclusions}

Analyzing the entire sample of galaxies where black hole mass, velocity dispersion $\sigma$ and asymptotic circular 
velocity $V_c$ have  all been measured,  we obtain the best-fit power law relation between \msigma\ and \mvc\ 
and find that the scatter and slope are very similar for both relations. This model-independent fit  suggests that the \mvc\ 
correlation is just as strong (or just as weak) as the correlation between \msigma\ given current
sample sizes. As noted by KB, the correlations worsen (or disappears) for $\sigma$ and $V_c$ outside a 180-260 \kms range. 

In this Letter, we have argued that in the context of our current understanding of the growth of galaxies and black hole assembly,  this is not unexpected.  Most importantly, the absence of central SMBHs in bulgeless low mass galaxies today is not necessarily an indication of the lack of correlation between the dark matter halo and the central object at every epoch. In fact, it is a consequence of the weaker coupling of the dark matter halo with central black holes for low mass halos that experience practically no major mergers in their lifetime.  
With an explicit example we show that black hole seed models at high redshift that do assume a strong coupling between SMBH seed masses and the spin of the dark matter halo, do predict that low mass bulgeless galaxies today should be bereft of central black holes. The key reason for this late time consequence is that SMBH formation is less efficient in low-mass galaxies with large angular momentum and these seeds are also less likely to grow.

Additionally, the growth of SMBHs in low-mass galaxies is hampered, as such galaxies rarely experience major mergers. In particular,  as suggested by KBC secular effects that can build pseudo-bulges might at late times decouple the properties of the central stellar-dominated region from the overall dark matter halo.  Besides, dynamical effects such as  gravitational recoil and ejection, increase deviations preferentially at $V_c$ below $\sim$100-200 km/s.  Therefore, we propose that dark matter halos do drive the overall formation and growth of black holes, in the sense that they set the stage for SMBH growth through the merger history of the host, set by large scale structure. However, especially for low-mass galaxies where the star formation history and SMBH growth are not driven by major mergers, we do not expect strong correlations between $\sigma$ (set by the stellar component) or $M_{\rm BH}$ \citep[set by baryonic processes such as accretion of gas released by stellar winds, see][]{Volonteri2011} and $V_c$ (set by the dark matter halo) today.

On the issue of central SMBHs in cluster galaxies, we note that tidal stripping strongly
truncates the dark matter distribution of galaxies infalling into clusters, while leaving the inner regions
that are primarily baryonic unaltered (see Natarajan et al. 2009).  Besides, it appears that tidal
stripping is more efficient for late-type galaxies. Therefore, the relation between
$\sigma$ and $V_c$  for late-type cluster galaxies (more so than for early-types) will not be the 
same as for equivalent luminosity field galaxies. Regardless of which is the more fundamental 
correlation, $M_{\rm BH}-\sigma$ or  $M_{\rm BH}-V_c$, these correlations are likely to differ
more significantly for cluster galaxies.

\begin{acknowledgments}

PN thanks the John Simon Guggenheim Foundation for the award of a Guggenheim fellowship and
the Institute for Theory and Computation at the Harvard Smithsonian Center for Astrophysics for hosting
her. MV acknowledges support from SAO Award TM1-12007X and NASA awards ATP NNX10AC84G and NNX07AH22G.

\end{acknowledgments}

\end{document}